\documentclass[11pt]{article}
\usepackage{moriond,epsfig}
\usepackage{wrapfig}
\usepackage{ amssymb }
\def\Journal#1#2#3#4{{#1} {\bf #2}, #3 (#4)}

\def\NPB{{\em Nucl. Phys.} B}
\def\PLB{{\em Phys. Lett.}  B}

\usepackage{url}

\def\be{\begin{equation}}
\def\ee{\end{equation}}
\def\bea{\begin{eqnarray}}
\def\eea{\end{eqnarray}}

\begin{document}
\vspace*{4cm}
\title{Search for R-hadrons at the ATLAS experiment at the LHC}

\author{M.D. Joergensen\\ On behalf of the ATLAS Collaboration}

\address{Niels Bohr Institute, University of Copenhagen\\ Blegdamsvej 17, Copenhagen, Denmark}

\maketitle\abstracts{
The latest search for massive long-lived hadronising particles with the ATLAS detector is presented. The search is conducted with the inner detector and an integrated luminosity corresponding to $2.06 \ \mathrm{fb}^{-1}$ at $\sqrt{s} = 7$ TeV. For Split-SUSY scenarios gluino masses below $810$ GeV have been excluded.}

\def\dedx{\mathrm{d}E / \mathrm{d}x}

\section{Introduction} 
\label{sec:introduction}
This paper presents the latest search for massive long-lived hadronising particles with the ATLAS detector at the LHC~\cite{conf22}.
The discovery of new massive long-lived particles at the LHC would be of fundamental significance~\cite{milstead}. It was the discovery of a stable massive particle that opened the field of particle physics, and for many years electrons, protons, neutrons and muons were the only known type of elementary particles, all of which could be detected due to their long-lived nature. With the advent of accelerators the discovery of long-lived hadrons heralded the development of quantum chromodynamics, the theory of the strong interaction. In that sense history is a primary motivation for the search for new long-lived particles. Furthermore multiple new models predicts potentially long-lived particles. The search presented here investigates specific SUSY scenarios that predict massive long-lived coloured particles in the form of colour-octet gluinos ($\tilde{g}$). Due to colour confinement it is expected that the heavy sparton will hadronise with standard model partons into a colour neutral object called an $R$-hadron~\cite{milstead}. In $R$-parity conserving SUSY models this bound state is kept stable. The sparton itself is assumed stable or at least long-lived by other mechanisms, for instance requiring a highly suppressed decay channel~\cite{ArkaniHamed:2004yi}.

The primary search strategy for these objects relies on them being massive compared to the available collision energy. This in turn leads to the assumption that the particles must move at low velocities compared to the speed of light, thus behaving differently than most other directly observable particles produced. If the slow moving particles carry electric charge the particle can be detected in multiple sub systems in ATLAS~\cite{tech}. In general two complementary methods both sensitive to particle velocity; time of flight and anomalous energy deposition are used. 

In previous searches~\cite{Aad20111,Aad20114} ATLAS reported results based on a combination of sub-detectors, Pixel based specific energy loss $\dedx$, Calorimeter based time-of-flight (ToF) and ToF in the Muon spectrometer as well. The analysis presented here only uses the Pixel based $\dedx$ method in order to remain insensitive to hadronic interactions in the calorimeters that can cause an exchange of light quarks making the $R$-hadron electrically neutral and invisible to further detection.



\section{Models and datasets} 
\label{sec:models_and_datasets}

The analysis is based on LHC $p p$ collision data at $\sqrt{s} = 7 \ \mathrm{TeV}$, recorded in the period between March and August 2011. This corresponds to an to an integrated luminosity of $2.06 \ \mathrm{fb}^{-1}$ after data quality cuts.
Signal samples have been generated in a range of masses from $200$ to $1000$ GeV in $100$ GeV intervals. The analysis is insensitive to lower masses due to trigger constraints and large backgrounds.
The R-Hadron samples are pair produced $\tilde{g} \tilde{g}$ simulated with {\sc Pythia}~\cite{Sjostrand:2006za}. The initial hadronisation of gluinos into R-Hadrons is done with a custom version of the Lund string hadronisation model. The particles are always produced stable in this scenario.
Simulating the R-Hadron interaction with material in the detector is handled by dedicated {\sc Geant4} routines~\cite{Kraan:2004tz,Mackeprang:2006gx}. Because of the limited knowledge regarding the possible hadronic states, three different models are used to cover various model assumptions~\cite{Aad20111}.
In all Monte Carlo samples, the primary collision event is overlaid with minimum bias simulated events to model the pile-up conditions in data.



\section{Pixel-based mass estimation} 
\label{sec:pixel_based_mass_estimation}
%

The silicon pixel detector is the part of the tracker situated closest to the beam pipe in ATLAS and covers the radial distance $50.5 < r < 122.5 \ \mathrm{mm}$ in the pseudorapidity region $|\eta| < 2.5$. The detector consists of three layers parallel to the beam pipe in the barrel region and perpendicular in in the end-cap regions. In total the pixel tracker contains $80$ million pixels. The pixel tracker is capable of reading out a ``time over threshold'' (ToT) value that corresponds to the amount of charge collected in each pixel. The threshold is set at a minimum value of $3200\pm80 \ e^{-}$. The maximum charge collected per pixel is limited by the dynamic range of the read out electronics (8 bit ADC) to roughly $8.5$ times the energy deposited by a minimum ionising particle (MIP). Depositions higher than that causes a saturation leading to a mismeasurement. The maximum value limits the sensitively to slow particles with velocities higher than $\beta \gtrsim 0.3$. To arrive at an estimate of the specific energy loss $\dedx$ the charge deposition is measured in all pixels in the vicinity of the track in all layers. The truncated\footnote{This is done to avoid fluctuations from a deposition far out on the Landau tail} mean value of the energy deposited is then divided by the expected track length in the material. 

\begin{figure}[ht]
	\centering
\begin{minipage}[t]{0.35\linewidth}
\centering
\includegraphics[width=\linewidth]{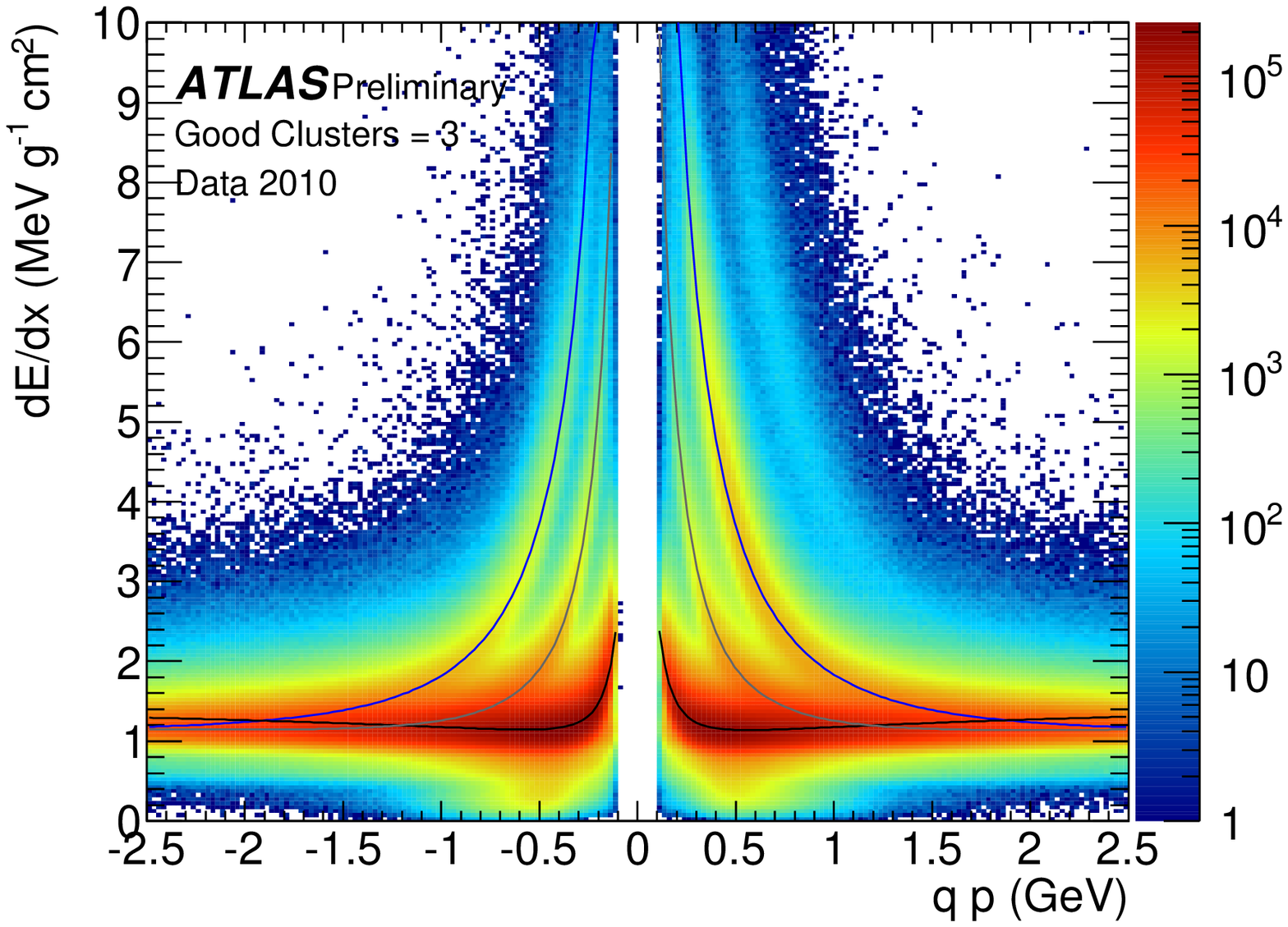}
\caption{Measured specific energy loss ($\dedx$) in the pixel detector as a function of momentum for low $p_T$ standard model particles~\protect\cite{conf22}. The bands corresponds to pions, kaons and protons respectively. }
\label{fig:dedx_p_sm}
\end{minipage}
\hspace{0.5cm}
\begin{minipage}[t]{0.35\linewidth}
\centering
\includegraphics[width=\linewidth]{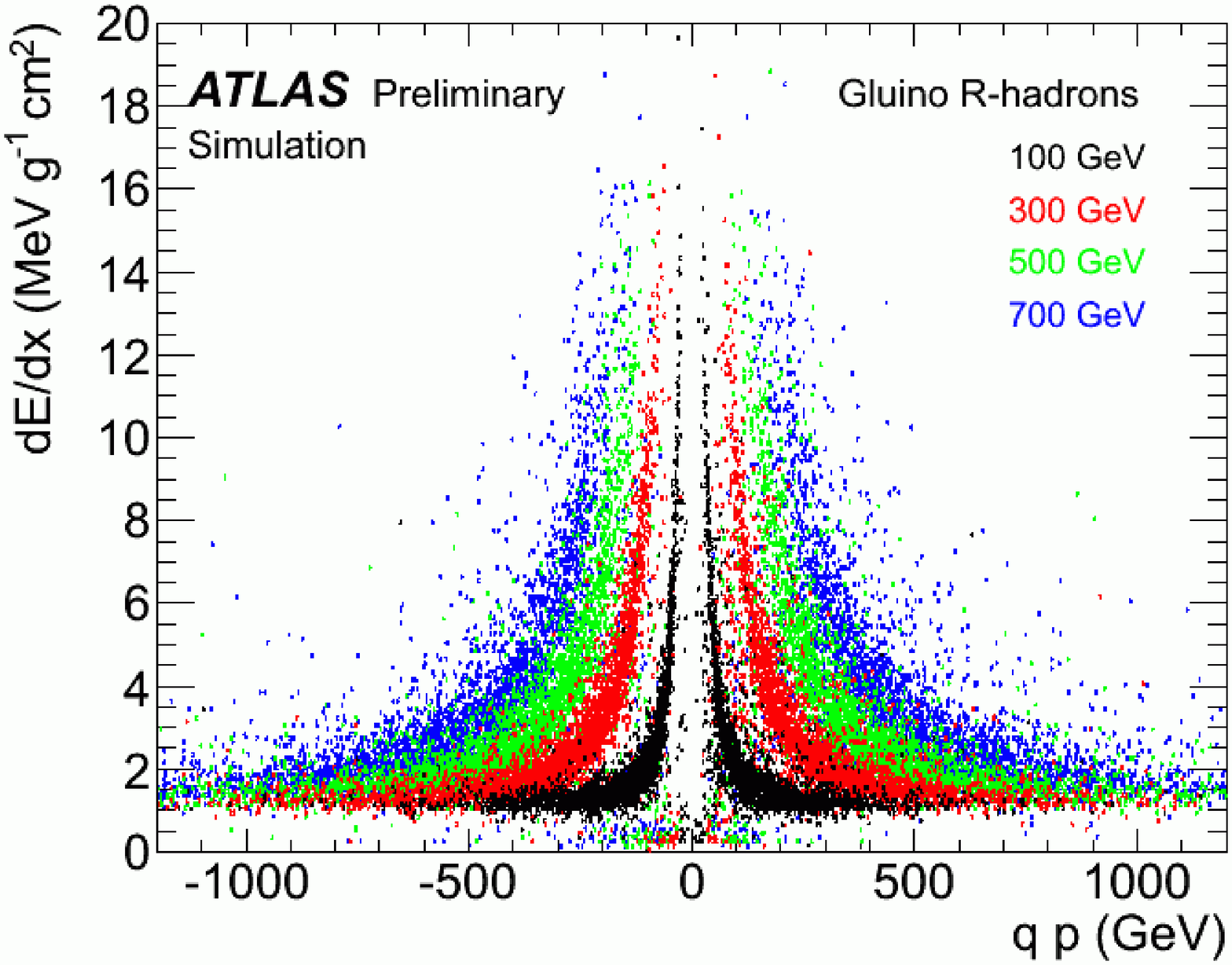}
\caption{Specific energy loss as a function of momentum for four different simulated R-hadron masses~\protect\cite{conf22}: 100, 300, 500 and 700 GeV.}
\label{fig:dedx_p_sig}
\end{minipage}
\end{figure}

The specific energy loss by a charged particle depends from the velocity of the particle in the material as illustrated by figures~\ref{fig:dedx_p_sm},~\ref{fig:dedx_p_sig}. The relationship is described by the well known Bethe-Bloch formula. To obtain an empirical formula for the specific energy loss in the pixel detector a fit is made on the $\dedx$ distributions from low momentum standard model particles with well-defined masses (fig.~\ref{fig:dedx_p_sm}).

Equation~\ref{eq:mpv} expresses the most probable value of the specific energy loss for a given velocity $\beta$. The function is numerically inverted to find the velocity as a function of energy loss and then the mass.
\begin{equation}
	\label{eq:mpv}
	\mathcal{MPV}_{\mathrm{d}E/\mathrm{d}x}(\beta) = \frac{p_1}{\beta^{p^3}} \ln \left(1+\left(p_2 \beta\gamma\right)^{p_5}\right) - p_4
\end{equation}



\section{Selection} 
\label{sec:selection}

\subsection{Trigger} 
\label{sub:trigger}
It is currently not possible to trigger directly on heavily ionising tracks with the ATLAS detector. To compensate, the analysis uses a missing energy trigger that relies on the production mechanism of the gluinos which also induces associated QCD radiation. A large missing energy component from initial state radiation (ISR) jets enables the use of the lowest unprescaled missing $E_T$ trigger available in the data taking period. The threshold is set to $E_T^{miss} > 70 \ \mathrm{GeV}$. The overall acceptance in the mass range of interest is around $20 \%$, with a slight mass dependence.

\subsection{Candidate selection} 
\label{sub:candidate_selection}

The selection reduces overall background by requiring offline $E_T^{miss} > 85 \ \mathrm{GeV}$, a vertex with at least $5$ tracks of which one track must have $p_T > 50 \ \mathrm{GeV}$ and the longitudinal and transverse impact parameters below $1.5$ mm. High-$p_T$ track candidates must also have three hits in the pixel detector, six hits in the silicon tracker and be isolated from other tracks with $p_T > 5$ GeV and $\Delta R_{trk} > 0.25$. Furthermore the track must have a momentum $p$ larger than $100$ GeV. The last cut is based on ionisation. The $\dedx$ estimate is not entirely flat as a function of pseudorapidity, to compensate the track must have a energy loss larger than $1.8 - 0.045 |\eta| + 0.115|\eta|^2 - 0.033|\eta|^3 \ \mathrm{MeV/g \ cm^{-2}}$. The resulting reduction is illustrated in table~\ref{tbl:cutflow} for data, background and two signal sets.

\begin{table}[t]
	\caption{Cut flow table: observed data yields and signal expectation~\protect\cite{conf22}. The quoted uncertainties are statistical.}
	\label{tbl:cutflow}
	\begin{tabular}{|l|rcl|c|c|c|}
		\hline
	& \multicolumn{3}{c|}{Data} & Bkg &  Gluino 400 GeV & Gluino 700 GeV \\
	\hline
	Cut level & \# Events    & Cut Eff. & Total Eff. 		   & \# Events & Total Eff. & Total Eff.       \\
	\hline                                                      
	Trigger   & 2,413,863   &         &        				   & --- & $0.205\pm0.013$ & $0.219\pm0.009$\\
	Offline $E_{T}^{miss}$   & $1,421,497$ & $0.589$& $0.589$  & --- & $0.200\pm0.013$ & $0.216\pm0.009$\\
	Primary vtx   & $1,368,821$ & $0.963$ & $0.567$ 		   & --- & $0.200\pm0.013$ & $0.216\pm0.009$\\
	High-$p_T$   & $212,464$ & $0.155$ & $0.0880$  			   & --- & $0.120\pm0.010$ & $0.129\pm0.007$\\
	Isolation  & $32,188$ & $0.151$ & $0.0133$ 				   & --- & $0.100\pm0.009$ & $0.105\pm0.006$\\
	High-$p$  & $21,040$ & $0.654$ & $8.7 \times 10^{-3}$	   & --- & $0.099\pm0.009$ & $0.104\pm0.006$\\
	Ionisation   & $\mathbf{333}$ & $0.016$ & $1.4 \times 10^{-4}$  	& $\mathbf{332}$ & $\mathbf{0.067\pm0.008}$ & $\mathbf{0.085\pm0.005}$\\
	\hline
	\end{tabular}
	
\end{table}
\subsection{Background estimation} 
\label{sub:background_estimation}
The background is expected to be entirely due to instrumental misclassification. A full Monte-Carlo simulation from first principles would require a very accurate model of the detector feedback to a degree not yet feasible. Instead a data-driven approach is used. By assuming independence between momentum and charge deposition for the background, two independent side-band windows (\textit{Bkg1} and \textit{Bkg2} in table~\ref{tbl:sidebands}) can be defined outside the relevant physics region.

\begin{table}[h]
	\label{tbl:sidebands}
	\caption{Sideband regions}
	\centering
	\begin{tabular}{|ccc|}
	\hline
	& Low $\dedx$ & High $\dedx$\\
	\hline
	Low p   & ---   &  Bkg2\\
	High p   & Bkg1  & signal \\
	\hline
	\end{tabular}
	\label{tbl:bgest}
\end{table}

In the \textit{Bkg1} region momentum is sampled as a function of $\eta$ with the requirement that $\dedx$ must be less than $1.8 \ \mathrm{MeV} \mathrm{g}^{-1} \mathrm{cm}^2$. 
In the \textit{Bkg2} region the $\dedx$ spectrum is sampled as a function of $\eta$ for $p < 100$ GeV.
Tracks in these regions are collected in 2D histograms $Bkg1(\eta,p)$, $Bkg2(\eta,\mathrm{d}E/\mathrm{d}x)$. The two distributions are then sampled randomly $2$ million times for each sampling and the values are combined to form a mass estimate based on the $\dedx$ and the momentum sampled. The resulting distribution is scaled to match the distribution in data by matching the sideband region below the signal region $m < 140$ GeV. The resulting background is shown in figure~\ref{fig:massspec}.



\section{Results} 
\label{sec:results}

\begin{figure}[ht]
\begin{minipage}[t]{0.5\linewidth}
\centering
\includegraphics[width=\linewidth]{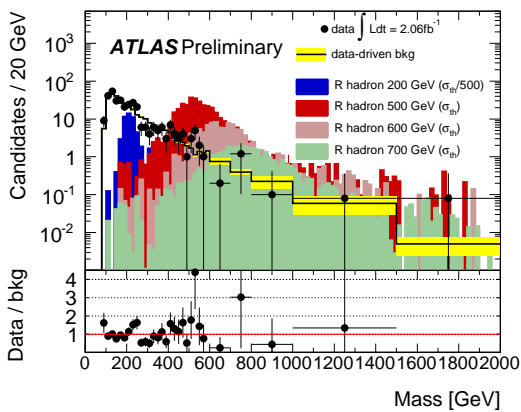}
\caption{Mass distribution for data, background and four simulated gluino R-Hadron signal scenarios scaled to the {\sc Prospino} expected cross sections~\protect\cite{conf22}.}
\label{fig:massspec}
\end{minipage}
\hspace{0.1cm}
\begin{minipage}[t]{0.5\linewidth}
\centering
\includegraphics[width=\linewidth]{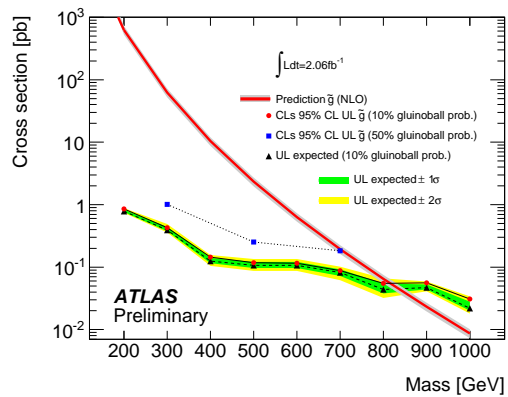}
\caption{Upper limit on gluino R-Hadron production cross section~\protect\cite{conf22}. }
\label{fig:limit}
\end{minipage}
\end{figure}

After the final selection no significant deviation from the standard model is observed. An upper limit is calculated for each mass hypothesis with the $\mathrm{CL_s}$ method at $95 \% \ \mathrm{CL}$. Comparing with expected cross sections produced with {\sc Prospino}~\cite{prospino} a mass exclusion for Split-SUSY gluinos are found for gluino masses less than $810$ GeV assuming gluino-gluon formation at $10 \%$. Limits for all tested masses are shown in figure~\ref{fig:limit}.

\section*{References}

\end{document}